\RequirePackage{snapshot}
\documentclass[conference,a4paper]{IEEEtran}

\usepackage{bm,bbm}
\usepackage{amsmath,amssymb}
\usepackage{graphicx,subfigure}
\usepackage{units}
\usepackage{url}
\usepackage{booktabs}
\usepackage{cite}
\usepackage{balance}

\begin{document}

\title{The Generalized Statistical Complexity of PolSAR Data}
\author{
\IEEEauthorblockN{Alejandro C.\ Frery$^1$, Eliana S.\ de Almeida$^1$, Osvaldo A.\ Rosso$^{1,2}$}
\IEEEauthorblockA{$^1$LaCCAN -- Laborat\'orio de Computa\c c\~ao Cient\'ifica e An\'alise Num\'erica\\
Universidade Federal de Alagoas\\
Av. Lourival Melo Mota, s/n\\
57072-900 Macei\'o -- AL, Brazil\\
\texttt{[acfrery;eliana.almeida]@gmail.com}\\
$^2$Laboratorio de Sistemas Complejos, Facultad de Ingenier\'{\i}a\\
Universidad de Buenos Aires\\
Av.\ Paseo Col\'on 840, Ciudad Aut\'onoma de Buenos Aires, 1063 Argentina\\
\texttt{oarosso@gmail.com}}}

\maketitle

\begin{abstract}
This paper presents and discusses the use of a new feature for PolSAR imagery: the Generalized Statistical Complexity.
This measure is able to capture the disorder of the data by means of the entropy, as well as its departure from a reference distribution.
The latter component is obtained by measuring a stochastic distance between two models: the $\mathcal G^0$ and the Gamma laws.
Preliminary results on the intensity components of AIRSAR image of San Francisco are encouraging.
\end{abstract}

\section{Introduction}

This paper discusses the use of the Generalized Statistical Complexity (GSC) as a feature for Polarimetric SAR (PolSAR) image analysis.

The GSC was proposed by Rosso et al.~\cite{GeneralizedStatisticalComplexityMeasure}.
It is an extension of the notion of order/disorder (uncertainty), which is conveniently captured by the entropy, to encompass the idea of structure, which is measured by a distance to an equilibrium distribution.
The GSC is the normalized product between an entropy and an stochastic distance.

Almeida et al.~\cite{CIARP2012GeneralizedStatisticalComplexitySARImagery} employed this idea to the analysis of intensity SAR data.
They used the Shannon entropy along with the Hellinger distance between a $\mathcal G^0$ and a gamma distribution.

In this paper we compute features (mean, scale, and texture) from the three intensity channels of a PolSAR image under the multiplicative model, and we show that the GSC provides additional information.

\section{The model and the feature}

Assuming the scaled Wishart distribution for full polarimetric observations, each pixel in a PolSAR image returns a complex positive definite random matrix 
$$
\bm Z = \left(
\begin{array}{ccc}
I_{11}	& A_{12} + j B_{12}	& A_{13} + j B_{13} \\
A_{12} - j B_{12}			& I_{22}						& A_{23} + j B_{23} \\
A_{13} - j B_{13}			&	A_{23} - j B_{23}							& I_{33}
\end{array}
\right),
$$ 
whose distribution is characterized by the density
\begin{equation}
 f_{\boldsymbol{Z}}(\boldsymbol{Z}';\boldsymbol{\Sigma},L) = \frac{L^{mL}|\boldsymbol{Z}'|^{L-m}}{|\boldsymbol{\Sigma}|^L \Gamma_m(L)} \exp\bigl\{
-L\operatorname{tr}\bigl(\boldsymbol{\Sigma}^{-1}\boldsymbol{Z}'\bigr)\bigr\},
\label{eq:Wishartdensity}
\end{equation}
where $m$ is the number of polarization channels, $\boldsymbol{\Sigma}$ is the complex covariance matrix of size $m\times m$, $L$ is the number of looks, $\Gamma_m(L)=\pi^{m(m-1)/2}\prod_{i=0}^{L-1}\Gamma(L-i)$ is the multivariate gamma function, and $|\cdot|$ and $\operatorname{tr}(\cdot)$ are the determinant and the trace, respectively.
With this, each intensity channel is described by the Gamma distribution~\cite{TriVariateChiSquaredFromWishart} with density given by
\begin{equation}
f_{Z_i}(Z'_i;L/\sigma^2_{i},L)=\frac{L^L{Z'_i}^{L-1}}{ \sigma^{2L}_i\Gamma(L)}
\exp\bigl\{-LZ'_i/\sigma^2_i\bigr\} \mathbbm 1_{\mathbbm R_+}(Z_i'), \label{eq:Gammadensity}
\end{equation}
for $i\in\{\text{11,22,33}\}$, where $\sigma^2_i$ is the $(i,i)$ entry of $\boldsymbol{\Sigma}$, and $Z'_i$ is the $(i,i)$ entry of the random matrix $\boldsymbol{Z}$. 

The scaled Wishart distribution is associated to fully developed speckle, i.e., there is no texture in the wavelength of the illumination due to the presence of infinitely many elements in the resolution cell, with each contributing infinitesimally to the return.
Freitas et al.~\cite{FreitasFreryCorreia:Environmetrics:03} proposed an extension for this model with nicer analytic properties than those of the Polarimetric K distribution~\cite{yueh89}.
Among them, the model for each the intensity channel is the $\mathcal G^0$ distribution with density
\begin{equation} 
f_{Z}(z;\alpha,\gamma,L)= \frac{L^L \Gamma{(L-\alpha)}}{\gamma^\alpha
\Gamma{(-\alpha)} \Gamma{(L)}} z^{L-1} \left(\gamma+Lz
\right)^{\alpha-L}  \mathbbm 1_{\mathbbm R_+}(z). \label{modelmultiplicative-22}
\end{equation}
The parameter $\alpha<0$ is a measure of texture, while $\gamma>0$ is proportional to the scale.
The maximum likelihood estimator for $(\alpha,\gamma)$, namely $(\widehat{\alpha},\widehat{\gamma})$, is the solution of the following system of non-linear equations:
\begin{align}
\psi^{0}(L-{\widehat\alpha})-\psi^{0}(-{\widehat\alpha})-\log \widehat\gamma+\frac1n\sum_{i=1}^n \log\left( \widehat\gamma+Lz_i\right)=0,\nonumber\\
-\frac{{\widehat\alpha}}{{\widehat\gamma}}+\frac{{\widehat\alpha}-L}{n}\sum_{i=1}^n({\widehat\gamma}+Lz_i)^{-1}=0,
\label{eq:mle}
 \end{align}
where $n$ is the sample size, and $\psi^{0}(\cdot)$ is the digamma function.

As proved by Frery et al.~\cite{frery96}, the connection between the $\mathcal G^0$ and Gamma distributions is provided by the following limit property: if $\alpha\to-\infty$ and $\gamma\to\infty$ such that $-\gamma/\alpha\to\sigma^2$ then the former becomes the latter, as characterized by equation~\eqref{eq:Gammadensity}.
This connection is the rationale behind the next proposal, namely, using the Gamma distribution as the equilibrium law in the computation of the GSC.

The information content of a system is typically described by the probability distribution of some measurable or observable quantity, and an information measure can be viewed as a quantity associated to this distribution. 
The Shannon entropy is often used as a the ``natural" one~\cite{Shannon1949}; it can be regarded as a measure of the uncertainty associated to the physical process described by the distribution.

Entropy measures do not quantify the degree of structure or patterns present in a process~\cite{Feldman1998}, which is not revealed by measures of randomness. 
The extremes of perfect order (like a periodic sequence) and of maximal randomness (fair coin toss) possess no complex structure and exhibit zero statistical complexity, with a range of possible degrees of physical structure between these extremes that should be quantified by {\it statistical complexity measures}. 
Rosso et al.~\cite{Lamberti2004} introduced an effective statistical complexity measure (SCM) that is able to detect essential details of the dynamics and differentiate different degrees of periodicity and chaos. 
This specific SCM, abbreviated as MPR, provides important additional information regarding the peculiarities of the underlying probability distribution, not already detected by the entropy.

The statistical complexity measure is defined,  following L\'opez-Ruiz  et al.~\cite{LopezRuiz1995}, via the product $C[P] =  H[P] \cdot D[P, P_{\text{ref}}]$, where $H$ is an entropy, $P$ is the distribution of the observed quantity, $D$ is a stochastic distance and $P_{\text{ref}}$ is a distribution of reference. 
The Statistical Complexity aims at measuring at the same time the order/disorder of the system by its entropy $H$, and how far the system is from its equilibrium state (the so-called disequilibrium $D$)~\cite{GeneralizedStatisticalComplexityMeasuresGeometricalAnalyticalProperties,GeneralizedStatisticalComplexityMeasure}.
In the case of PolSAR imagery, the equilibrium distribution is the Wishart law, since it describes fully developed speckle, i.e., situations where there is no texture.
Salicr\'u et al.~\cite{salicruetal1993,Salicru1994} provide a very convenient conceptual framework for both the entropy and the stochastic distance.

Let $f_{\boldsymbol{Z}}(\boldsymbol{Z}';\boldsymbol{\theta})$ be a probability density function with parameter vector $\boldsymbol{\theta}$ which characterizes the distribution of the (possibly multivariate) random variable $\boldsymbol{Z}$.
The ($h,\phi$)-entropy relative to $\boldsymbol{Z}$ is defined by 
\begin{align*}
H_{\phi}^h(\boldsymbol{\theta})=h\Big(\int_{\mathcal A}\phi(f_{\boldsymbol{Z}}(\boldsymbol{Z}';\boldsymbol{\theta}))\mathrm{d}\boldsymbol{Z}'\Big),
\end{align*}
where either $\phi:\bigl[0,\infty\bigr) \rightarrow \mathbb{R}$ is concave and $h:\mathbb{R} \rightarrow \mathbb{R}$ is increasing, or $\phi$ is convex and $h$ is decreasing.   
The differential element $\mathrm{d}\boldsymbol{Z}'$ sweeps the whole support $\mathcal A$.
In this work we only employ the Shannon entropy, for which $h(y)=y$ and $\phi(x)=-x\ln x$.

Consider now the (possibly multivariate) random variables $\boldsymbol{X}$ and $\boldsymbol{Y}$ with densities $f_{\boldsymbol{X}}(Z;\boldsymbol{\theta_1})$ and $f_{\boldsymbol{Y}}(Z;\boldsymbol{\theta_2})$, respectively, where $\boldsymbol{\theta_1}$ and $\boldsymbol{\theta_2}$ are parameter vectors.
The densities are assumed to have the same support $\boldsymbol{\mathcal A}$.
The $(h,\phi)$-divergence between $f_{\boldsymbol{X}}$ and $f_{\boldsymbol{Y}}$ is defined by
\begin{equation} 
D_{\phi}^h(\boldsymbol{X},\boldsymbol{Y}) = 
h\biggl(\int_{\boldsymbol{\mathcal A}} \phi\biggl( \frac{f_{\boldsymbol{X}}({Z};\boldsymbol{\theta_1})}{f_{\boldsymbol{Y}}({Z};\boldsymbol{\theta_2})}\biggr) f_{\boldsymbol{Y}}({Z};\boldsymbol{\theta_2})\mathrm{d}{Z}\biggr),
\label{eq:eps2-no}
\end{equation}
where $h\colon(0,\infty)\rightarrow[0,\infty)$ is a strictly increasing function with $h(0)=0$ and $\phi\colon (0,\infty)\rightarrow[0,\infty)$ is a convex function such that $0\,\phi(0/0)=0$ and $0\,\phi(x/0)=\lim_{x \rightarrow \,\infty} \phi(x)/x$. 

Following Almeida et al.~\cite{CIARP2012GeneralizedStatisticalComplexitySARImagery}, we will only employ the Hellinger divergence which is also a distance, for which $h(y)={y}/{2}$, $0\leq y<2$ and $\phi(x)=(\sqrt{x}-1)^2$, and we define the Statistical Complexity of coordinate $(i,j)$ in an intensity SAR image as the product
\begin{equation}
C(i,j) = H(i,j) D(i,j), \label{eq:StatisticalComplexity}
\end{equation}
where $H(i,j)$ is the Shannon entropy observed in $(i,j)$ under the $\mathcal G^0$ model, and $D(i,j)$ is the observed Hellinger distance between the universal model (the $\mathcal G^0$ distribution) and the reference model of fully developed speckle (the $\Gamma$ law).

The Hellinger distance and the Shannon entropy were computed numerically.
We define the PolSAR GSC as the vector-valued operator which returns the GSC of each intensity channel.

\section{Results}

We used the National Aeronautics and Space Administration Jet Propulsion Laboratory (NASA/JPL) Airborne SAR (AIRSAR) image of the San Francisco Bay, obtained in the \textsf{L}-band, with four nominal looks, and \unit[$10\times10$]{m$^2$} of spatial resolution.

In a first approach, samples of size $101\times101$ of the main three classes (sea, forest and urban) were extracted and analyzed from the HH channel.
The mean $\sigma$ which indexes the Gamma law, the parameters of the $\mathcal G^0$ distribution, the Hellinger distance between these two models, the entropy under the $\mathcal G^0$ law and, finally, the Generalized Statistical Complexity of these three representative samples are presented in Table~\ref{tab:ThreeSamples}.

\begin{table}[hbt]
\centering
\caption{Estimated quantities in large homogeneous samples}\label{tab:ThreeSamples}
\begin{tabular}{c *{3}{c}} \toprule
 & Sea & Forest & Urban\\ \midrule
$\widehat\sigma$ 										& $0.0294$ 				& $0.0983$ 					& $0.1670$\\
$(\widehat{\alpha}, \widehat{\gamma})$ 	& $(-11.870,0.320)$	& $(-2.717,  0.179)$	& $(-2.051,  0.182)$\\
$H$ 																& $2.790$ 					& $1.400$						& $0.928$\\
$D$ 																& $0.0066$ 				& $0.0669$					& $0.110$\\
$C$ 																& $0.0184$				& $0.0936$					& $0.102$\\ \bottomrule
\end{tabular}
\end{table}

As expected, the texture parameter $\alpha$ increases with the roughness of the sample but, in this image, the difference between forest and urban areas is not particularly strong.
The entropy follows the opposite behavior, since it reduces when the texture increases.
The distance between the Gamma and the $\mathcal G^0$ models exhibits the expected behavior: in areas with little or no texture, the roughness parameter is small and, therefore, the models tend to coincide, i.e., the Gamma distribution is a good descriptor for the data, as well as the more general $\mathcal G^0$ law.
When the texture increases, the Gamma model looses its ability to follow the data, and it progressively yields worse and worse fits than the $\mathcal G^0$ distribution, leading to increased distances between them.
The complexity, being the product of the entropy and the distance to the reference model, is able to detect the difference between textured and non-textured areas and, to a a lesser extent, the difference between degrees of texture.

\begin{figure*}[hbt]
\centering
\subfigure[Sea data]{\includegraphics[width=.28\linewidth]{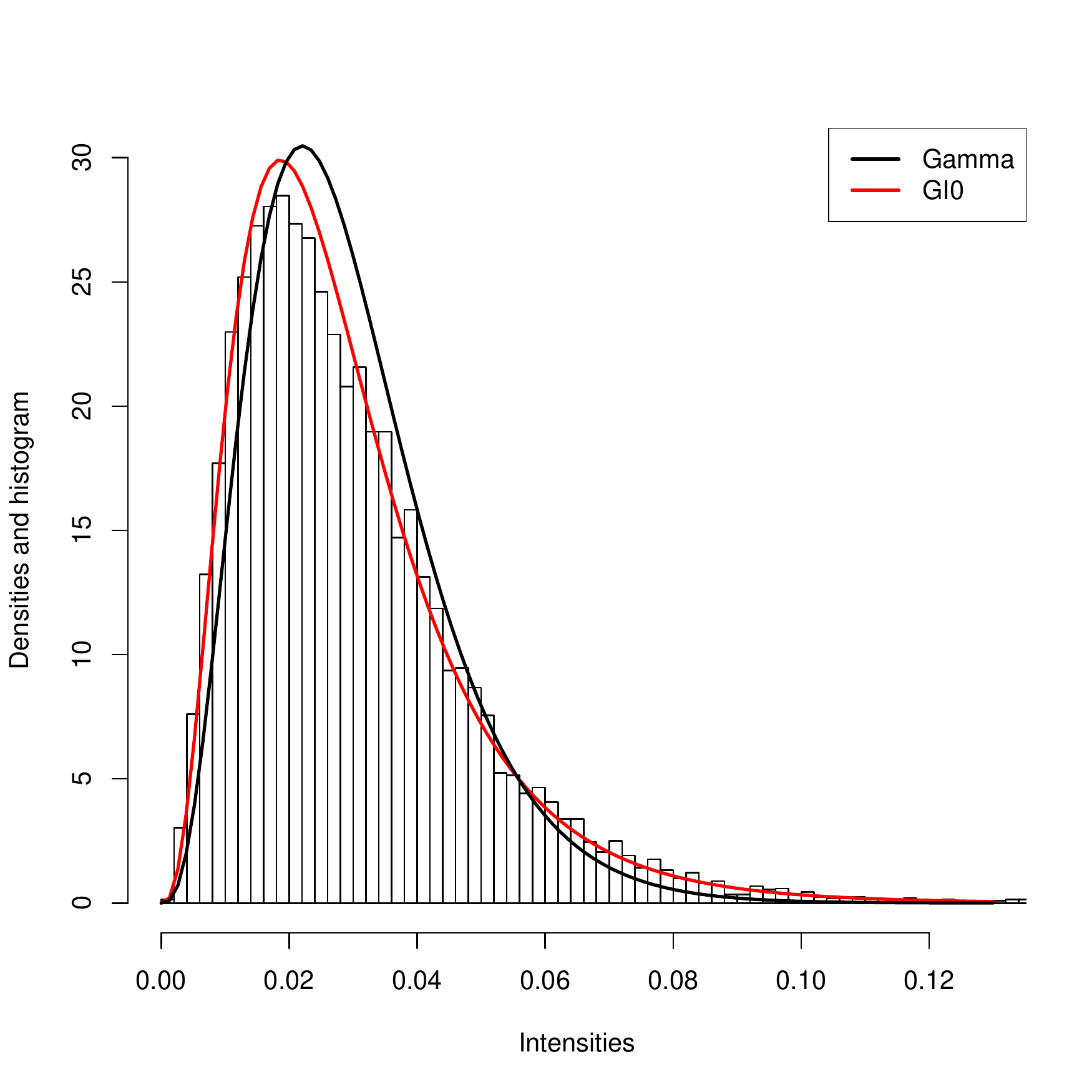}}
\subfigure[Forest data]{\includegraphics[width=.28\linewidth]{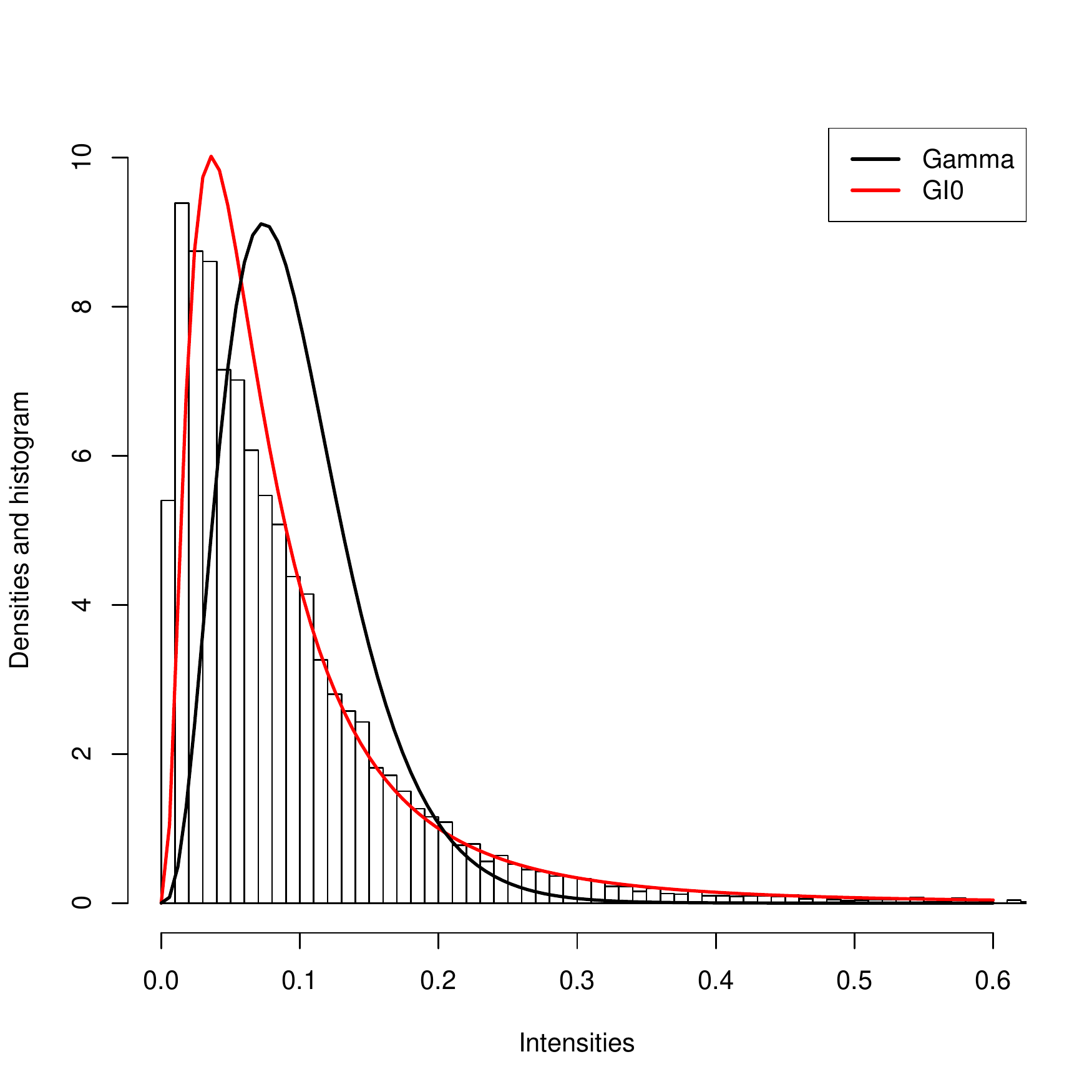}}
\subfigure[Urban data]{\includegraphics[width=.28\linewidth]{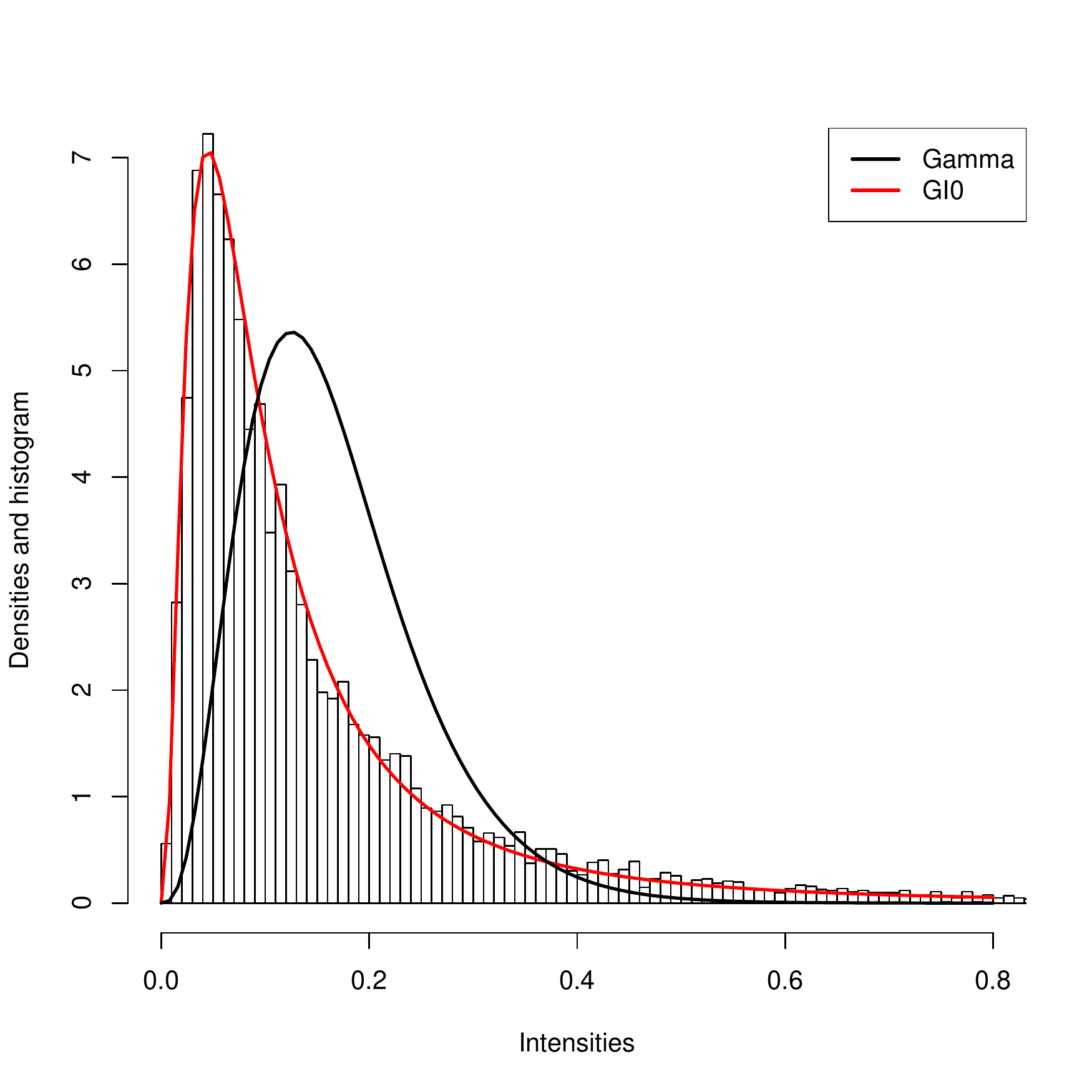}}
\caption{Histograms and fitted models}\label{fig:ThreeSamplesFit}
\end{figure*}

Figure~\ref{fig:ThreeSamplesFit} presents the histograms and the two fitted models for each sample.
The $\mathcal G^0$ model describes all types of samples with excellent expresiveness, deserving the denomination ``Universal Model'' proposed by Mejail et al.~\cite{MejailJacoboFreryBustos:IJRS}.
The ability of the Gamma distribution to describe the data is limited to the textureless sample, i.e., to the sea.
The more textured the target, the worse the fit provided by the reference model.

The densities presented in Figure~\ref{fig:ThreeSamplesFit} justify the results exhibited in Table~\ref{tab:ThreeSamples} regarding the stochastic distances.
Homogeneous samples lead to reduced distances, since both the Gamma and the $\mathcal G^0$ distribution are good models, and they agree producing very similar fits.
The densities mostly overlap, leading to distances close to zero.
In the case of extreme texture, the Gamma law is unable to capture the variabilty of the data, while the $\mathcal G^0$ model provides a very good fit.
This leads to very different densities, which are further apart with respect to the stochastic distance here considered.

Once verified the ability of the measures here proposed to capture the features of interest, we move on to extract these features locally in every image coordinate.

We computed the following features in windows of size $7\times7$: the mean, estimators of $\alpha $ and $\gamma$ from the $\mathcal G^0$ distribution, Shannon entropy, Hellinger distance between the best fit of the $\mathcal G^0$ and $\Gamma$ laws, and the GSC.

\begin{figure*}[hbt]
\centering
\subfigure[Local mean\label{fig:mean}]{\includegraphics[width=.32\linewidth]{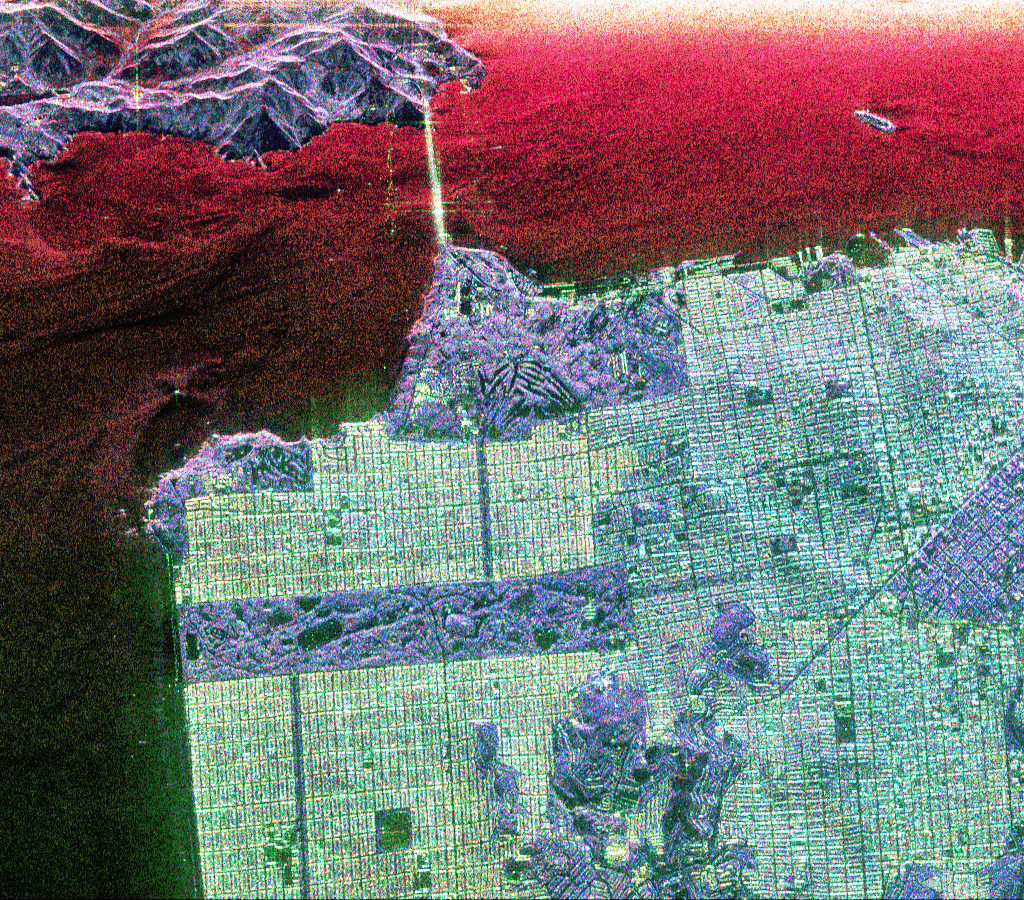}}
\subfigure[Texture estimate $\widehat{\alpha}$\label{fig:AlphaEstimate}]{\includegraphics[width=.32\linewidth]{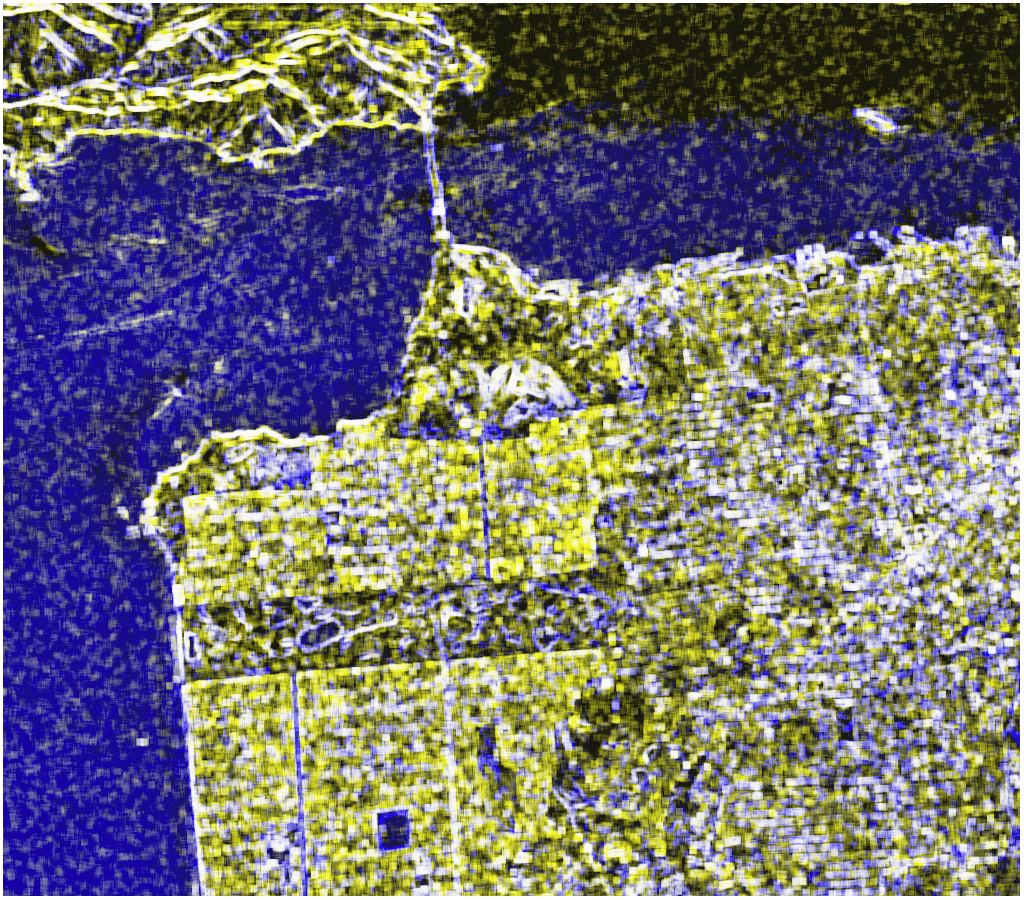}}
\subfigure[Scale estimate $\widehat{\gamma}$\label{fig:GammaEstimate}]{\includegraphics[width=.32\linewidth]{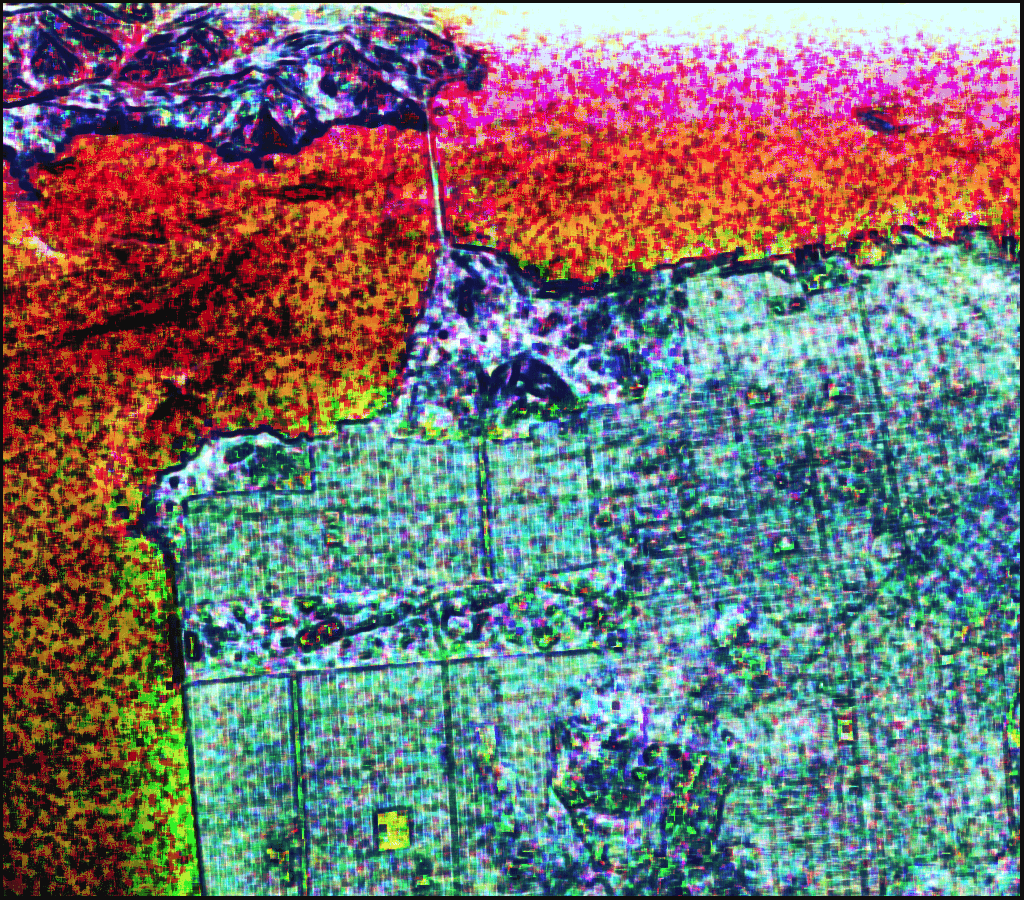}}
\subfigure[Shannon Entropy $H$\label{fig:Entropy}]{\includegraphics[width=.32\linewidth]{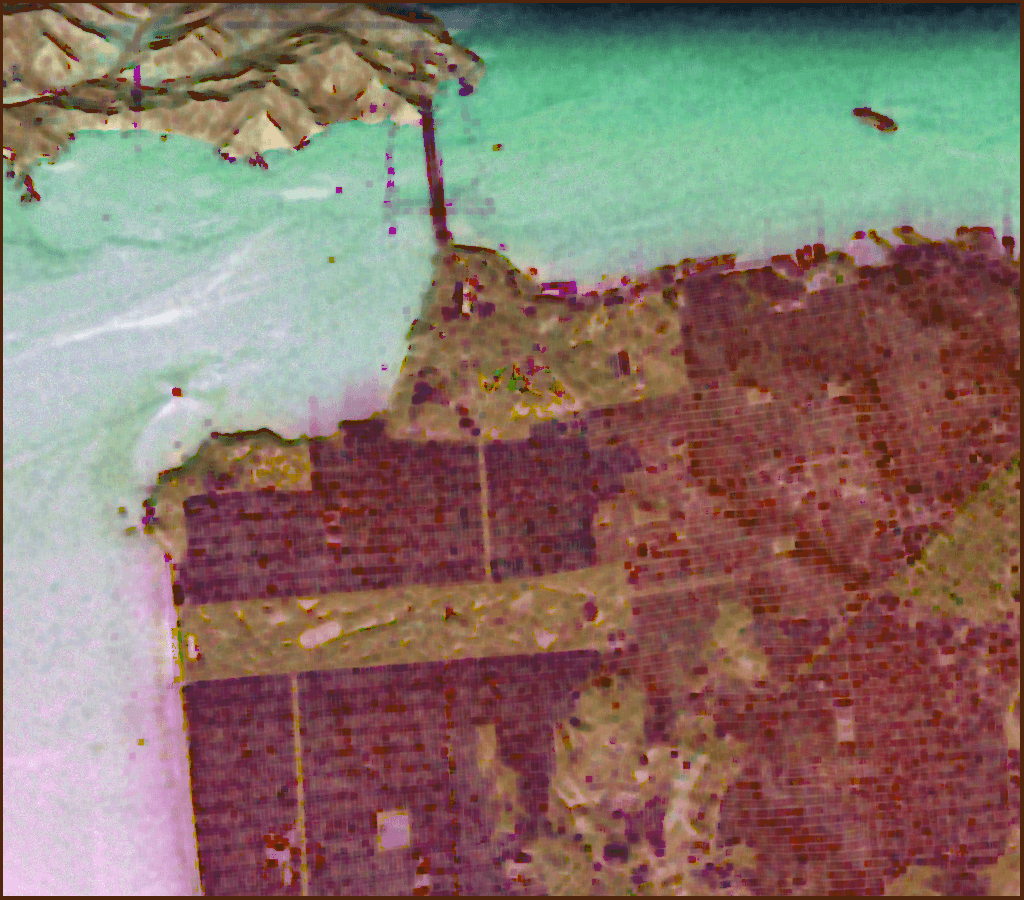}}
\subfigure[Hellinger distance $D$\label{fig:HellingerDistance}]{\includegraphics[width=.32\linewidth]{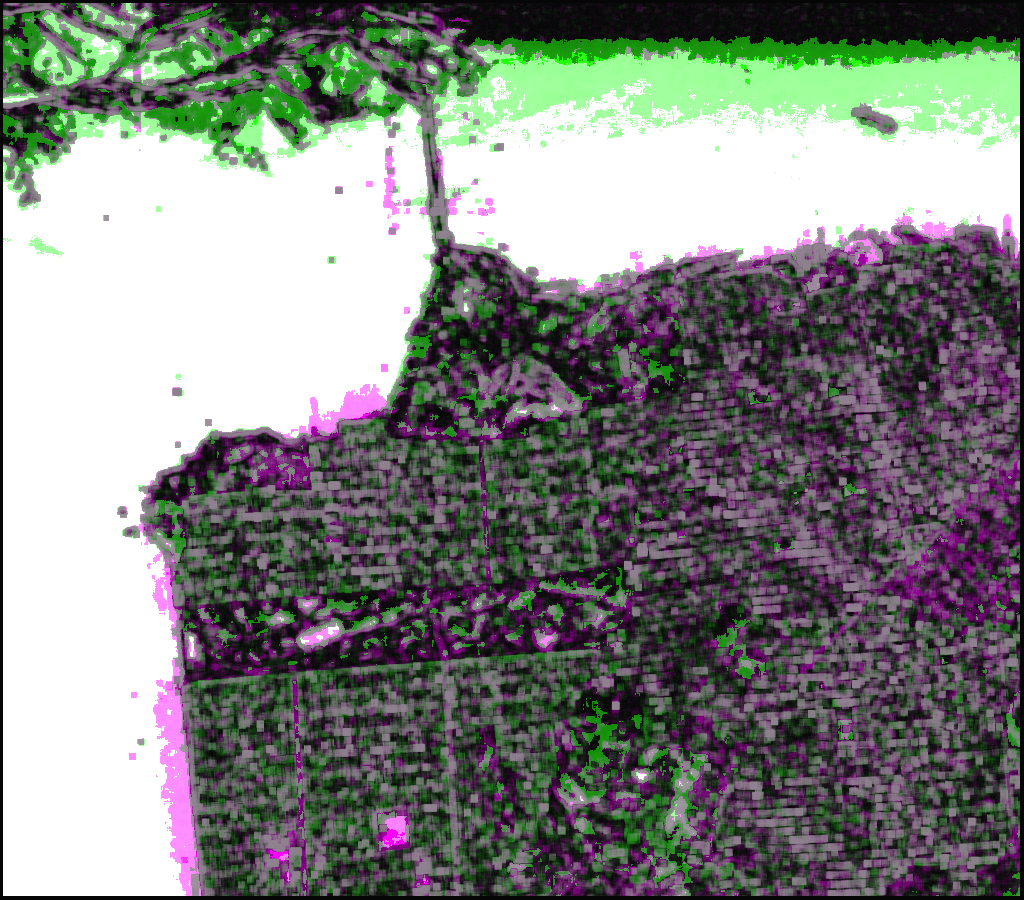}}
\subfigure[Generalized Statistical Complexity $C$\label{fig:GSC}]{\includegraphics[width=.32\linewidth]{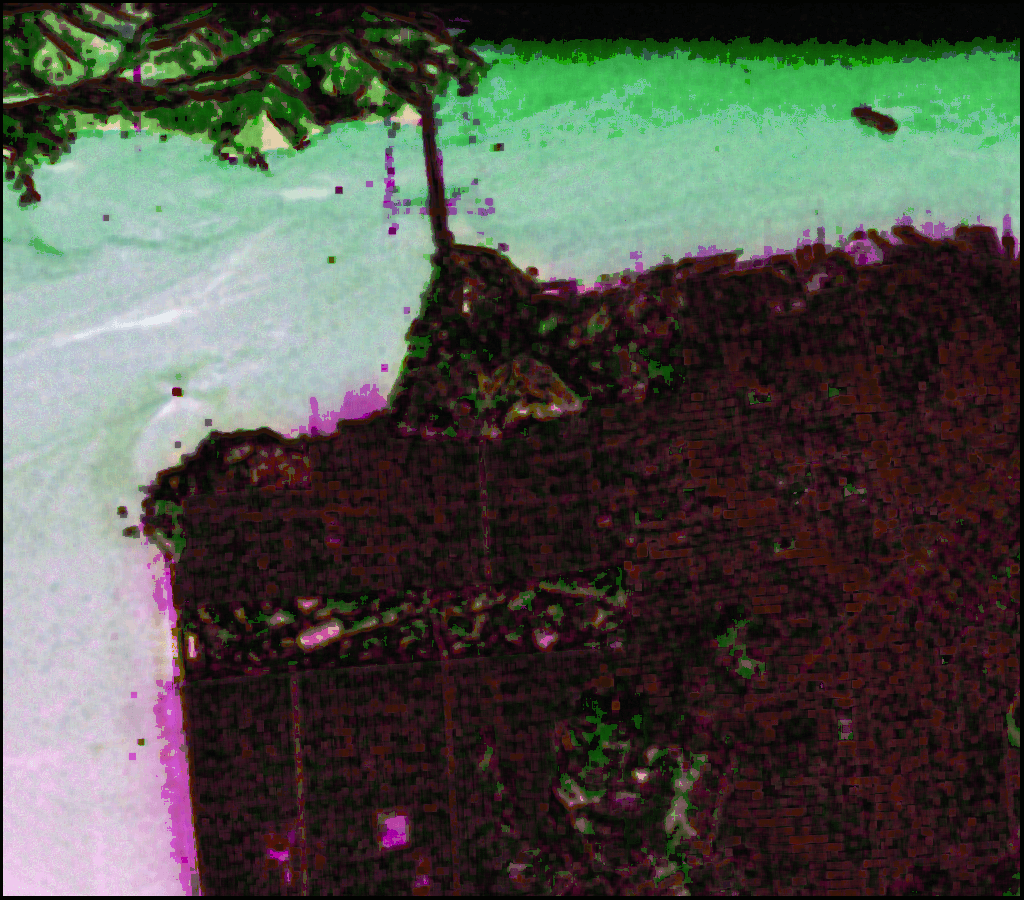}}
\caption{Features extracted from the PolSAR data.}\label{fig:SanFran}
\end{figure*}

These measures were computed on each channel.
Each feature extracted in the HH, HV and VV channels was equalized and then mapped to the Red, Green and Blue components to form false color images.
Figure~\ref{fig:SanFran} shows the main results.

Figure~\ref{fig:mean} shows the mean values; the urban areas and man-made structures stand brighter than forest which, in turn, is brighter than sea.
Figure~\ref{fig:AlphaEstimate} presents the texture estimates.
Although the sea is the less textured area, there are areas with higher return; there are waves in these areas which appear in light blue shades; this behavior will be more evident when computing the entropy and the distance.
Figure~\ref{fig:GammaEstimate} presents the scale estimates.

Figure~\ref{fig:Entropy} shows the entropy.
It is noticeable how this feature is able to retrieve the main classes, and delineates them with great precision.
A few spots in the sea may be the result of specular return.
Figure~\ref{fig:HellingerDistance} shows the Hellinger distances between the fully developed speckle model and the distribution which captures texture.
The high return from the sea tends to dominate this feature (which is shown after image equalization).
Nevertheless, the distance captures well the ground targets, and identifies correctly the urban spots and the park areas.
Notice that the urban area to the right of the image lies closer to the textureless model than the middle right; this is probably due to the fact that the former has more trees than the latter.

Figure~\ref{fig:GSC} shows the final result: the Generalized Statistical Complexity.
It identifies with great detail the linear features which correspond to roads, and other characteristics which are not so clear in the other features.

\section{Discussion}

The texture parameter $\alpha$, as discussed in previous works, is able to capture the target roughness, as seen in Figure~\ref{fig:AlphaEstimate}.
This information is valuable for identifying regions which only differ by their texture as, for instance, within the sea class.

The color composition of Shannon entropies (Fig.~\ref{fig:Entropy}) clearly distinguishes many types of targets, yielding an interesting feature for other procedures as, for instance, classification.

The Hellinger distances (Fig.~\ref{fig:HellingerDistance}) can be interpreted as an smoothed version of the texture parameter in inverse scale.

An investigation using field data is needed in order to identify the ground features which yield the different types of complexity.

The procedure requires the use of dependable estimates of the parameters of the $\mathcal G^0$ distribution, a subject which is still matter of research.
Two requirements are conflicting: on the one hand, the larger the sample the more precise the estimation, but also the more prone it will be to contamination from other classes; on the other hand, the smaller the sample, the more immune it will be to data from more than one class but, also, the less dependable the estimator will be in terms of bias, variance, and numerical stability.

Analytic expressions for the distance between the Gamma and the $\mathcal G^0$ model, as well as for the Shannon entropy under the latter are under assessment.

\balance
\bibliographystyle{IEEEtran}
\bibliography{art89,art93,art94,art95,art98,art04,art06,art10,art11,rdt,livros}
\end{document}